\documentclass[preprint,12pt]{elsarticle}

\usepackage{amssymb}

\usepackage{graphicx}
\usepackage{subfigure}
\usepackage{dcolumn}
\usepackage{bm}
\usepackage{enumitem}
\usepackage{amsmath,mathrsfs,amsfonts,dsfont,bm}
\usepackage{hyperref}
\hypersetup{colorlinks=true, citecolor=blue, urlcolor=blue, linkcolor=blue}
\journal{Physica A: Statistical Mechanics and its Applications}
\begin{document}

\begin{frontmatter}


\title{Social hierarchy promotes the cooperation prevalence}

\author[1]{Rizhou Liang}
\author[2,3]{Jiqiang Zhang}
\author[1]{Guozhong Zheng}
\author[1,4]{Li Chen}
\address[1]{School of Physics and Information Technology, Shaanxi Normal University, Xi'an 710061, P. R. China}
\address[2]{School of Physics and Electronic-Electrical Engineering, Ningxia University, Yinchuan 750021, P. R. China}
\address[3]{Beijing Advanced Innovation Center for Big Data and Brain Computing, Beihang University, Beijing 100191, P. R. China}
\address[4]{Email address: chenl@snnu.edu.cn}
\begin{abstract}
Social hierarchy is important that can not be ignored in human socioeconomic activities and in the animal world. Here we incorporate this factor into the evolutionary game to see what impact it could have on the cooperation outcome.  The probabilistic strategy adoption between two players is then not only determined by their payoffs, but also by their hierarchy difference  --- players in the high rank are more likely to reproduce their strategies than the peers in the low rank. Through simulating the evolution of Prisoners' dilemma game with three hierarchical distributions, we find that the levels of cooperation are enhanced in all cases, and the enhancement is optimal in the uniform case. The enhancement is due to the fact that the presence of hierarchy facilitates the formation of cooperation clusters with high-rank players acting as the nucleation cores. This mechanism remains valid on Barab\'asi-Albert scale-free networks, in particular the cooperation enhancement is maximal when the hubs are of higher social ranks. We also study a two-hierarchy model, where similar cooperation promotion is revealed and some theoretical analyses are provided. Our finding may partially explain why the social hierarchy is so ubiquitous on this planet.
\end{abstract}
\begin{keyword}
{social hierarchy \sep cooperation \sep complex networks}



\end{keyword}

\end{frontmatter}


\section{Introduction}
\label{}
From Confucius to Kant, cooperation is a central concern in our society not only for its humanity value but also its crucial role in the development of economy, technology, and science \emph{etc} ~\cite{Perc2016Phase}. Cooperation is also ubiquitous in the natural world, like those prosocial species, such as ants and bees, whose survivals rely on their altruistic behaviors~\cite{Worden2007Evolutionary}. But according to the evolution theory of Darwinism, individuals are inherently selfish therefore they are only willing to change their strategies or behaviors to maximize their profits. Because defection generally brings more profits as least in the short term, no one is willing to cooperate and the world will be dominated by the defectors. How the cooperation emerges and is maintained in selfish population is then a vital question that has attracted the attention of researchers in different fields in the past several decades.

In this regard, the evolutionary game theory provides a proper framework to explore the potential mechanisms behind cooperation ~\cite{Szabo2007Evolutionary,Nowak2004Evolutionary}, and some prototypical game models are often adopted for this aim, such as the prisoner's dilemma~\cite{Szab1998Evolutionary}, the snowdrift game~\cite{Shu2018Memory,Hauert2004Spatial} and the public good game~\cite{Archetti2012Review} \emph{etc}. The endeavor has successfully revealed quite a few mechanisms, such as direct reciprocity~\cite{Pacheo2008Repeated}, indirect reciprocity ~\cite{Nowak2005Evolution,Ohtsuki2006The,Ghang2015Indirect}, kin selection~\cite{Griffin2002Kin}, group selection, spatial or network reciprocity ~\cite{Nowak1992Evolutionary,Santos2005Scale-Free}, voluntary participation~\cite{Hauert2002Volunteering,Szabo2002Phase,Wu2005Spatial} and punishment ~\cite{Chen2015,Yang2018Promoting}. More recent advance comes from the methodologies within machine learning to compare what outcome difference could be if the games are played by AI algorithms~\cite{carleo2019machine,zhang2020oscillatory,zhang2020understanding}.
 Note that, similar to the spirit of identical particle assumption in statistical mechanics, individuals in all these models are supposed to be
indistinguishable and thus of equal position in the strategy updating.

However, a consensus is that our society is more often composed by heterogeneous individuals where they differ in many aspects such as their occupations, social statuses, cultural backgrounds \emph{etc}~\cite{Pride2010Pride} that potentially affect their decision-makings. 
In this regard, researchers have studied a bunch of related factors and revealed that these social diversities generally lead to better cooperation outcomes.
When the diversity is attributed to the number of interactions~\cite{Santos2005Scale-Free,Santos2008Social}, heterogeneous graphs such as scale-free networks are found to promote cooperation.
In ~\cite{Perce2008Social}, the social diversity is reflected in different mapping of game payoffs to individual fitness, and they find that the levels of cooperation are improved in general cases. Within the same spirit, this type of diversity is also able to promote cooperation in spatial multigames~\cite{Qin2017Social} where half of the population play prisoner's dilemma and the other half play the snowdrift game simultaneously. 
In ~\cite{Chen2018Popularity,Zhang2014Cooperation}, the researchers unveil that the popularity-driven selection generally facilitates the formation of cooperator clusters and thus cooperation, where the popularity, the degree of strategy homophily, can also be seen as a kind of social diversity. 
In a heterogeneously structured population, optimal cooperation also requires the diversity to be conformists, the masses conform but not the hubs~\cite{Szolnoki2016Leaders}.
The diversity of teaching or learning ability is also found to be influential; while the level of cooperation always promotes for diversified teaching, the impact of learning activity depends on the strategy updating scheme~\cite{Szolnoki2007Cooperation,Szolnoki2008Coevolution,Szolnoki2008Diversity,Wu2015Diverse}. 
Resource heterogeneity as another type of social diversity can also facilitate cooperation if it is well managed~\cite{KunResource}.

While the diversity emphasizes the inequality of individuals and has been extensively studied as listed above, another closely related but different concept is the hierarchy that instead focuses on the difference between individuals, which is relatively less touched. In fact, the hierarchy is so ubiquitous in almost all socialites that it deeply shaped the creature's behaviors and determined their survival~\cite{hammerstein2003genetic}.
Some evidences have been identified, indicating that the social hierarchy is beneficial to their survival in general. For example, in condor's group, dominance hierarchies could regulate competitive access to food resources through the direction of the hierarchy~\cite{Sheppard2013Hierarchical}. Also the dominance rank among wild female African elephants confers fitness benefits on the population because it improves the access to resources~\cite{Archie2006Dominance}. Moreover, the hierarchy in ants' society regulates the reproduction for efficient use of limited resources~\cite{Monnin1999Reproduction}. While the influence of social hierarchy on some other behaviors, e.g. collective motions, have been studied~\cite{nagy2010hierarchical,flack2018local,xue2020hierarchical}, it remains unclear that how the social hierarchy could affect the cooperation prevalence in general.

\begin{figure*}[htbp]
\includegraphics[width=1.0\linewidth]{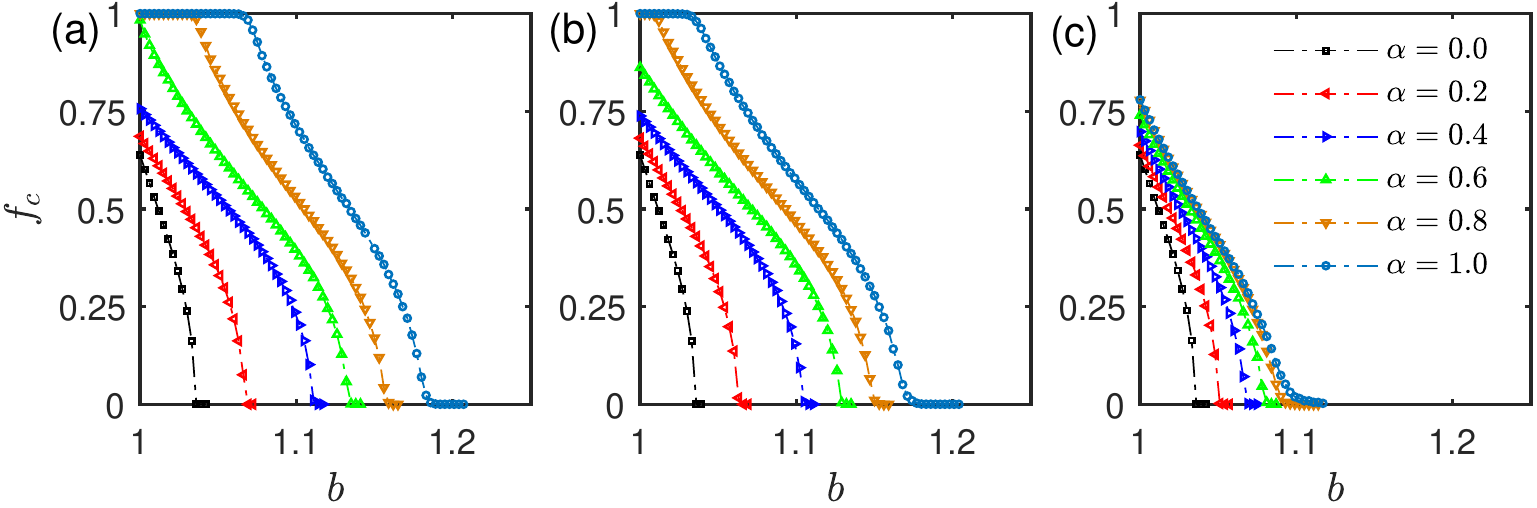}
\caption{
(Color online) Cooperator fraction $f_c$ as a function of temptation $b$ for a couple of hierarchy coefficients $\alpha$ given three different hierarchical distributions: (a) uniform, (b) exponential, and (c) power-law. The model returns to the original PD game model when $\alpha$=0. As $\alpha$ increases, the level of cooperation continuously improves. Other parameters: $L=1024$, $K=0.025$.
}
\label{fig:fc_PT}
\end{figure*}
Here, we introduce and investigate an evolutionary game model of a hierarchical population in Sec.~\ref{sec:model}, where each player is designated a social rank according to some distributions. For players in the high social rank, their strategies are more likely to reproduce in their neighborhood than players in the low rank, even if their payoffs are close. Specifically, we study the evolution of prisoner's dilemma game with three different hierarchy distributions, and we find that the levels of cooperation are significantly improved in all cases (Sec.~\ref{sec:results}). The promotion is due to the hierarchy-induced spatial structures that effectively protect the survival of cooperators. This promotion is observed in scale-free networks as well (Sec.~\ref{sec:sf}). We also develop a simplified model to give some analytic treatments (Sec.~\ref{sec:analysis}).

\section{Model}
\label{sec:model}
In our model, the system is composed of $N$ individuals that are located on an $L\times L$ square lattice with a periodic boundary condition. Each player is initially set either as a cooperator or defector with equal probability. They play pairwise game defined by three scenarios. Mutual cooperation brings each the reward $R$, mutual defection yields the punishment $P$ for each, and mixed encounter gives the cooperator the sucker's payoff $S$ yet the temptation $T$ for the defector. Strict prisoner's dilemma (PD) requires $T>R>P>S$, but here we adopt the common practice with parameterization $R=1$ , $P=S=0$, and $T=b$, which is known as the weak PD game. To incorporate the hierarchy, each player $i$ is designated a random hierarchical number $h_i\in[0,1)$ at beginning, drawn from some distributions. Players with higher value of $h$ are supposed to be in the higher social rank.

In an elementary step of the standard Monte Carlo (MC) simulation, the procedure is as follows. First, an individual $i$ is randomly chosen and acquires its mean payoff $\bar{\Pi}_i$ (defined by the total payoff $\Pi_i$ divided by its degree) by playing the game with all its neighbors $\Omega_{i}$ defined by the underlying networks. Next, one of $i$'s neighbors $j$ is selected randomly and also acquires its mean payoff $\bar{\Pi}_j$ by playing the game within its neighborhood $\Omega_{j}$. Lastly, player $i$ adopts the strategy of $j$ with an imitation probability according to the Fermi rule~\cite{Szab1998Evolutionary}
 \begin{equation}\label{eq:imitation}
 \begin{aligned}
 W(s_{j}\rightarrow s_{i})=\dfrac{1}{1+\exp[(\bar{\Pi}_{i}-\bar{\Pi}_{j}(1+\alpha\bigtriangleup h))/K]},
 \end{aligned}
\end{equation}
where $\Delta h=h_{j}-h_{i}\in(-1, 1)$ denotes the hierarchal difference between $i$ and $j$. $\alpha\in [0, 1]$ is the hierarchal coefficient determining the degree to which the strategy adoption in the imitation process is influenced by the hierarchy. Obviously, the case of $\alpha=0$ is recovered to the traditional evolutionary games, where the strategy updating is purely determined by their payoffs, irrespective of other factors. Instead, when $\alpha>0$, the presence of hierarchical difference facilitates the strategy reproduction of those players in the high social rank. $K$ quantifies the uncertainty in decision-making during the imitation, and is fixed at 0.025 throughout the whole study. A full MC step consists of $N$ such elementary steps, which means that every player is going to update its strategy once on average.

To be specific, we consider the following three distributions that were first proposed in~\cite{Perce2008Social}: uniform ($\propto const.$), exponential ($\propto e^{-2h}$), and power-law types ($\propto h^{-1}$), which are drawn in practice as follows:
\begin{equation}
\begin{aligned}
h &=\chi,\quad \chi\in[0,1), \\
h &=\dfrac{1}{2}\ln(\dfrac{1}{1-\chi}),\quad \chi\in[0,1-\dfrac{1}{e^{2}}),\\
h &=(\dfrac{1}{\epsilon})^{\chi-1}, \quad \chi\in[0,1).
\end{aligned}
\end{equation}
Here $\chi$ are uniformly drawn in the given range, and $h\in(\epsilon,1)$ with $\epsilon\rightarrow 0$ for the last implementation. For most numerical experiments, 50 thousand MC steps are run to guarantee that the equilibrium is reached, and then we average the data for another 10 thousand MC steps.

\begin{figure*}[htbp]
\subfigure{
\begin{minipage}[t]{0.3\textwidth}
\centering
\includegraphics[width=1.0\linewidth]{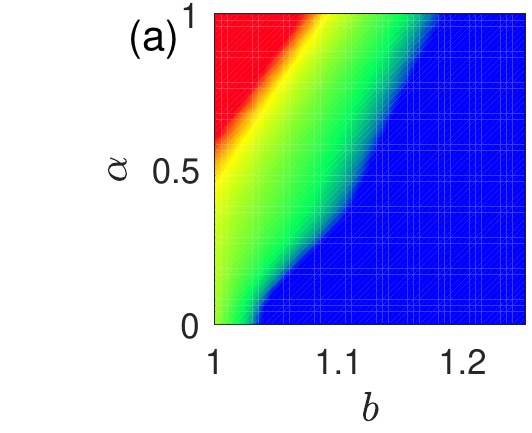}
\end{minipage}
}
\subfigure{
\begin{minipage}[t]{0.3\textwidth}
\centering
\includegraphics[width=1.0\linewidth]{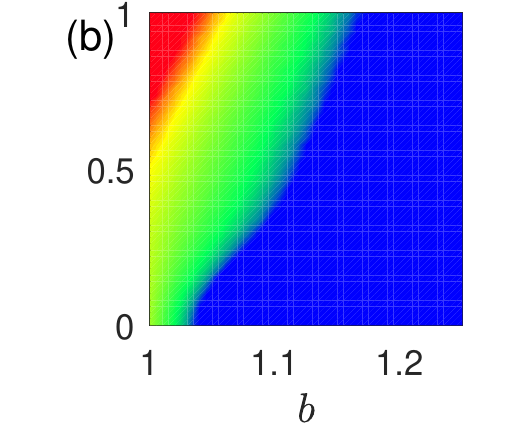}
\end{minipage}
}
\subfigure{
\begin{minipage}[t]{0.3\textwidth}
\centering
\includegraphics[width=1.0\linewidth]{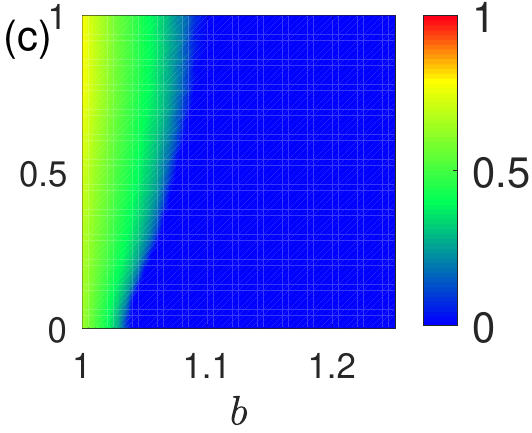}
\end{minipage}
}
\caption{
(Color online) Encoded color map for the cooperator fraction $f_c$ on the $b-\alpha$ parameter plane respectively for three different types of hierarchical distributions: (a) uniform, (b) exponential, and (c) power-law. Other parameters: $L=256$, $K=0.025$.
}
\label{fig:phase_diagram}
\end{figure*}

\section{Results}
\label{sec:results}
\subsection{Numerical simulations}

To begin with, we first present the impact of hierarchy on the cooperation prevalence in the uniform distribution case. Fig.~\ref{fig:fc_PT}(a) gives the cooperation fraction as the function of the temptation $b$ for a couple of hierarchical coefficients $\alpha$. It shows that as the impact of hierarchy becomes stronger, the cooperative likelihood is increased. Specifically, the threshold for cooperation outbreak $b_{c_1}$ (below which cooperators start to appear) is continuously increased as $\alpha$ becomes larger and this shift is the most significant for $\alpha=1$. Meanwhile, the threshold for defector eradication $b_{c_2}$ (below which defectors go extinct) is also increased, whereby full cooperation is possible in those strong hierarchy cases for the given parameter region. For the temptation $b_{c_2}<b<b_{c_1}$, the coexistence states of cooperators and defectors are expected.

For comparison, Fig.~\ref{fig:fc_PT}(b) and \ref{fig:fc_PT}(c) illustrate two cases with nonuniform distributions, which show qualitatively the same cooperation enhancement. Compared to the case of uniform distribution, the enhancement is lesser in the population with exponential and power-law hierarchical distributions, especially the cooperation promotion is least in the latter scenario, where the defector eradication threshold $b_{c_2}$ is not even present in the shown parameter range. But still the cooperation outbreak thresholds $b_{c_1}$ are shifted to the right and all prevalences are greater than the case without hierarchical impact ($\alpha=0$). Note that, in previous diversity studies~\cite{Perce2008Social,Qin2017Social}, the power-law distribution is shown to be optimal for the cooperation promotion, followed by the exponential distribution, the uniform distribution is the worst case. This is just opposite to our observations here.

To more systematically investigate the impact of hierarchy, Fig.~\ref{fig:phase_diagram} provides the phase diagrams in the $b-\alpha$ parameter space for the three distributions. It confirms that the cooperation is promoted in all cases as the hierarchy coefficient $\alpha$ becomes larger. In particular, both thresholds $b_{c_{1,2}}$ are monotonously shifted and they are almost linear functions of the hierarchical coefficient $\alpha$. These results suggest that the enhancement of cooperation in the hierarchical population is universal, and the uniform distribution of social ranking comparatively yields the optimal promotion of cooperation.

\subsection{Mechanism analysis}
To explore the mechanism behind the promotion, let's recall the basic supporting mechanism -- network reciprocity~\cite{Nowak1992Evolutionary}. Typically, when cooperators and defectors are randomly placed on the lattice, a decay of cooperation is expected at beginning because defectors jeopardise the reproduction of cooperation. But later on, cooperators are able to form cooperator clusters whereby they support each other and resist against the invasion of defectors at boundaries. This suggests that the presence of hierarchy could better enhance the formation of cooperation clusters than the traditional case ($\alpha=0$). The spatiotemporal evolution indeed confirms that cooperation clusters survive better when the hierarchy is incorporated (data not shown).

\begin{figure*}[htbp]
\includegraphics[width=1.0\linewidth]{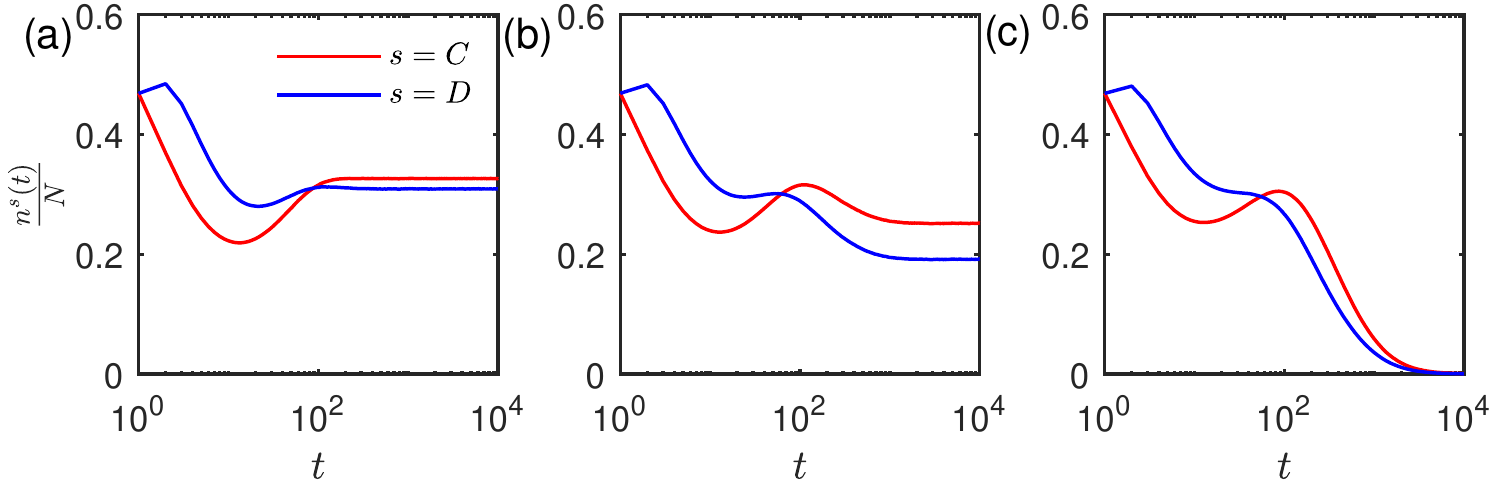}
\caption{(Color online) The fraction of all cooperators and defectors at boundaries for three hierarchical strengths $\alpha=0.5, 0.75, 1$. At very beginning, almost all players are at boundaries when starting from random initial conditions. By evolution, the boundary fractions decay and saturate in the case of $\alpha=0.5$ (a), but show nontrivial peaks in $\alpha=0.75$ (b) and $\alpha=1$ (c). Parameters: $L=1024$, $K=0.025$, and $b=1.05$.}
\label{fig:dis_total}
\end{figure*}

To understand how the hierarchy facilitates the formation of cooperation clusters, it's crucial to investigate the evolution dynamics at the interaction boundaries between the cooperator and defector clusters.
By the boundary, it is defined as the time-varying set $\mathcal B(t)$ of any site with at least one different state in its four nearest neighbors.
First, let's monitor the time evolution of the fraction of boundaries in the population defined as $n^s(t)/N$, shown in Fig.~\ref{fig:dis_total} for three typical hierarchical cases ($\alpha\!=\!0.5,0.75,1$). The rapid decrease in the first few steps in all cases is due to reduction of cooperators starting from the well-mixed condition, followed by a typical increase due to the formation and growth of cooperation clusters. While the increase saturates in the case of $\alpha=0.5$, there is a nontrivial peak in the cases of $\alpha=0.75$ and 1, especially for boundary cooperators. This reason lies in the fact that in these cases, many small cooperator clusters further merge into each other gradually, resulting in bigger ones which then reduce the boundaries as well as cooperators or defectors there.

To study the individual difference caused by the hierarchical rank, we classify all individuals into five subgroups on the basis of the hierarchical labeling as $L_g$ with $g=1,2,...,5$ if $h_i\in[0,0.2),[0.2,0.4),...,[0.8,1)$ respectively. Specifically, we monitor the evolution of relative composition with respect to the hierarchy for both cooperators and defectors at boundaries, and the relative composition fractions are defined as
 \begin{equation}\label{eq:definition}
 \begin{aligned}
 f^s_{L_g}=\dfrac{n^s_{L_g}(t)}{n^s(t)}, s\in{\{C,D\}}.
 \end{aligned}
\end{equation}
Here, $n^s_{L_g}(t)=\sum_{i\in \mathcal B_{L_g}(t)} \delta(s_i(t)-s)$ is the number of players within the state $s$ at the interface belonging to subgroup $L_g$ at time $t$, and $n^s(t)=\sum_{g=1}^5{n^s_{L_g}(t)}$ accordingly.

\begin{figure*}[htbp]
\includegraphics[width=1.0\linewidth]{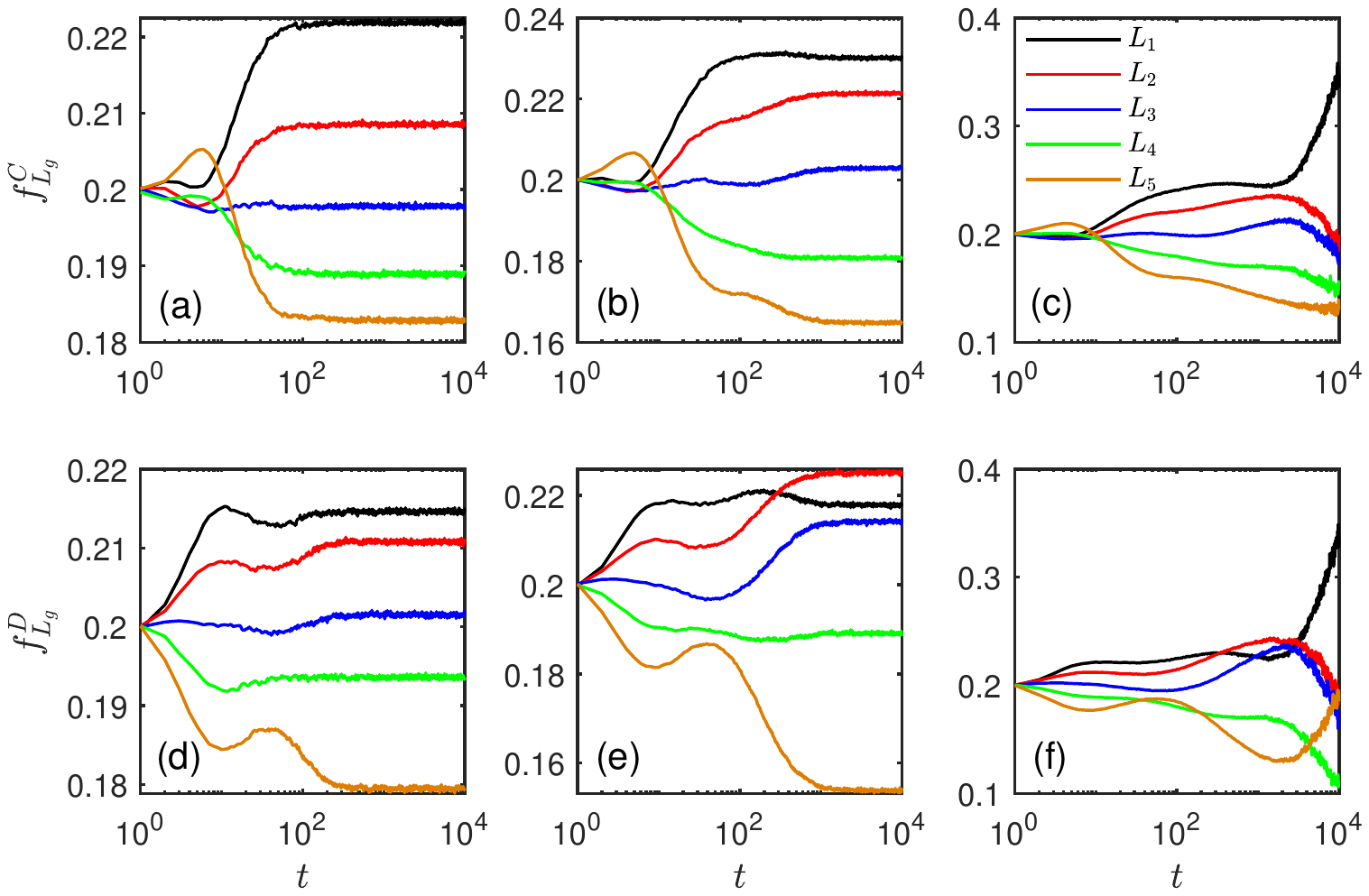}
\caption{(Color online) The time series of relative composition fraction at boundaries $f^s_{L_g}(t)$ for three hierarchical strengths $\alpha=0.5, 0.75, 1$. (a-c) and (d-f) are respectively for cooperators and defectors. In each subplot, the population is divided into five subgroups according to their hierarchy, $L_{1,2,...,5}$ correspond to the lowest to the highest ranks. Other parameters: $L=1024$, $K=0.025$, and $b=1.05$.
}
\label{fig:dis_CD}
\end{figure*}

Fig.~\ref{fig:dis_CD} further shows the relative hierarchical compositions $f^{C,D}_{L_g}$ respectively for the five subgroups. For boundary cooperators, it shows that at the very early stage only the relative fraction of the highest level of cooperators $L_5$ exhibits obvious increase. This observation is understandable because for well-mixed population at this stage defectors are at relatively advantage position over cooperators, but the high rank for those cooperators compensates their disadvantageous competitiveness, therefore they survive better than those in lower subgroups. This explains increasing trend for those higher rank subgroups of cooperators. Interestingly, this trend is reversed as time goes by that cooperators of lower subgroups are dominating at boundaries and the fraction differences become larger as the hierarchical impact becomes stronger. This means that in the long term the cooperation clusters are more likely surrounded by low-rank cooperators; and accordingly those high-rank cooperators are more probably located within the center position of cooperation clusters. This is reasonable because those high-rank cooperators who survive better in the early phase naturally act as the nucleation core for the growth of cooperation clusters at the late stage.

\begin{figure*}[htbp]
\includegraphics[width=1.0\linewidth]{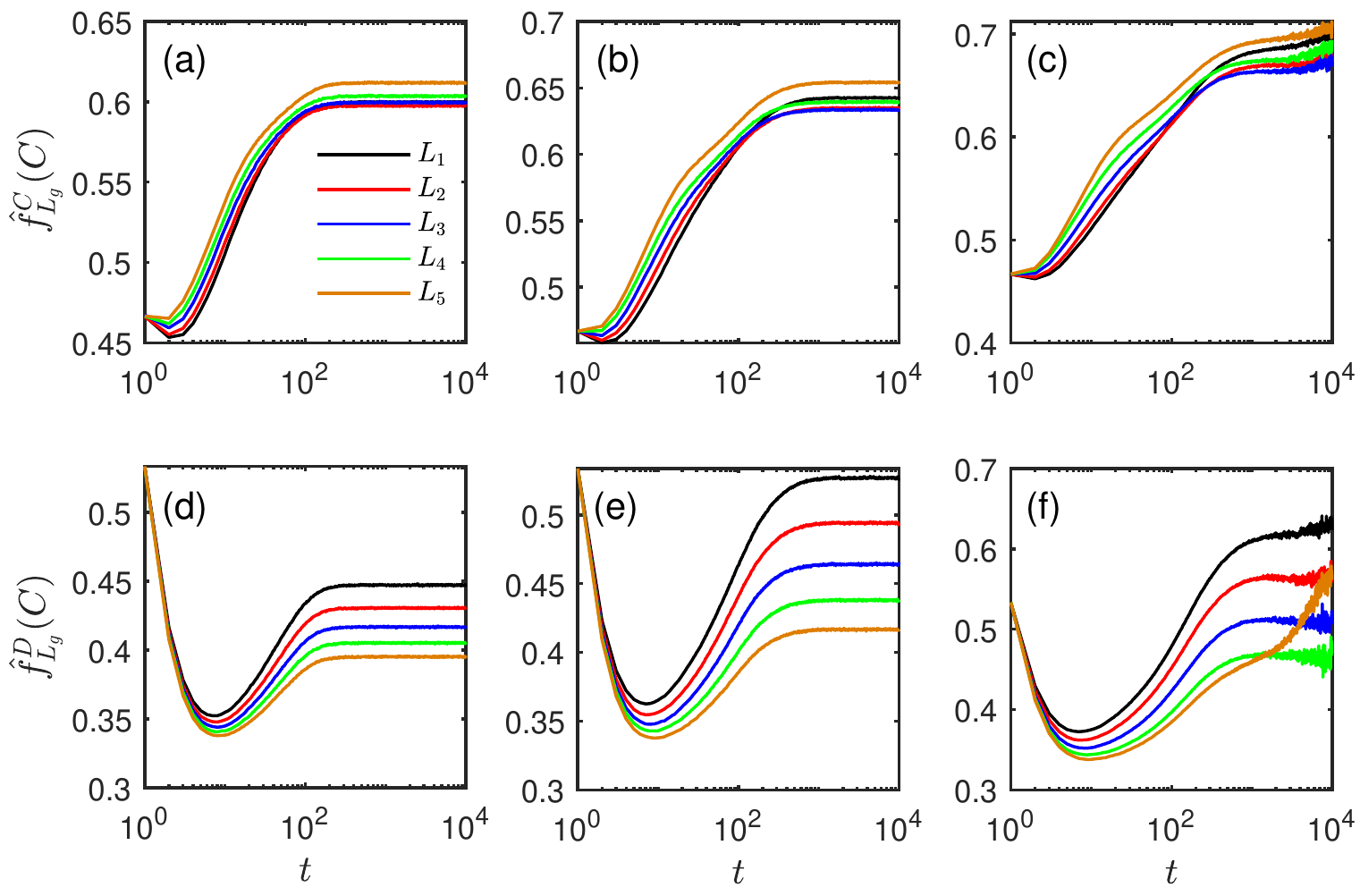}
\caption{(Color online) The time series of neighborhood composition fraction at boundaries $\hat{f}^s_{L_g}(t)$ for three hierarchical strengths $\alpha=0.5, 0.75, 1$. (a-c) and (d-f) are respectively for cooperators and defectors. Also, the population is divided into five subgroups. Other parameters: $L=1024$, $K=0.025$, and $b=1.05$.
}
\label{fig:dis_neighbors}
\end{figure*}

The time evolution of composition fractions for defectors, however, shows some different dynamical features, see the lower row in Fig.~\ref{fig:dis_CD}. The long term evolution shows qualitatively the same property that low-rank defectors dominate at the interface. But this feature evolves at the very beginning of evolution unlike the cooperator case. These means that high-rank defectors are not likely to appear at the interaction interface from the beginning. This difference lies in the fact that high-rank cooperators act as the core of cooperation clusters but those high-rank defectors are more often embedded in a connected defection sea.

To characterize the interaction interface in more details, we also survey the fractions of cooperators in their neighborhood centered around players at boundaries. According to both the state and hierarchy of the center players, we compute the evolution of cooperator fraction in the neighborhood respectively for cooperator and defector being the center player,
\begin{eqnarray}
\hat{f}_{L_g}^{s}(C)\!=\!\frac{\sum\limits_{i\in {\mathcal{B}_{L_g}}(t)}\sum\limits_{j\in\Omega_{i}}\delta(s_{i}-s)\delta(s_{j}-C)}
{4\sum\limits_{i\in {{\mathcal B}_{L_g}}(t)}\delta(s_{i}-s)}, s\!\in\!\{C, D\}
\end{eqnarray}
where the population is also divided into five subgroups. The result is shown in Fig.~\ref{fig:dis_neighbors}. In all cases, the cooperator fractions first decrease then followed by an increase and saturate in the end.
The most significant observation is that for cooperators at the boundary higher rank of their social hierarchy convinces more of their neighbors to be cooperative as well, and vice versa, while for defectors the opposite is true that higher ranks lead to much less cooperators in their neighborhood.

Altogether, these observations constitute the following picture: due to the presence of social hierarchy, those high-rank cooperators survive from exploitation starting from random conditions, and they act as the nucleation cores whereby cooperation clusters grow by attracting more and more low rank individuals around; at interaction boundaries high-rank cooperators facilitate the growth of cooperation clusters while high-rank defectors do the opposite. Without social hierarchy, the nucleation process is absent in the cases when the temptation $b$ is large, thus the cooperation cannot be expected.

\begin{figure}
\includegraphics[width=0.5\linewidth]{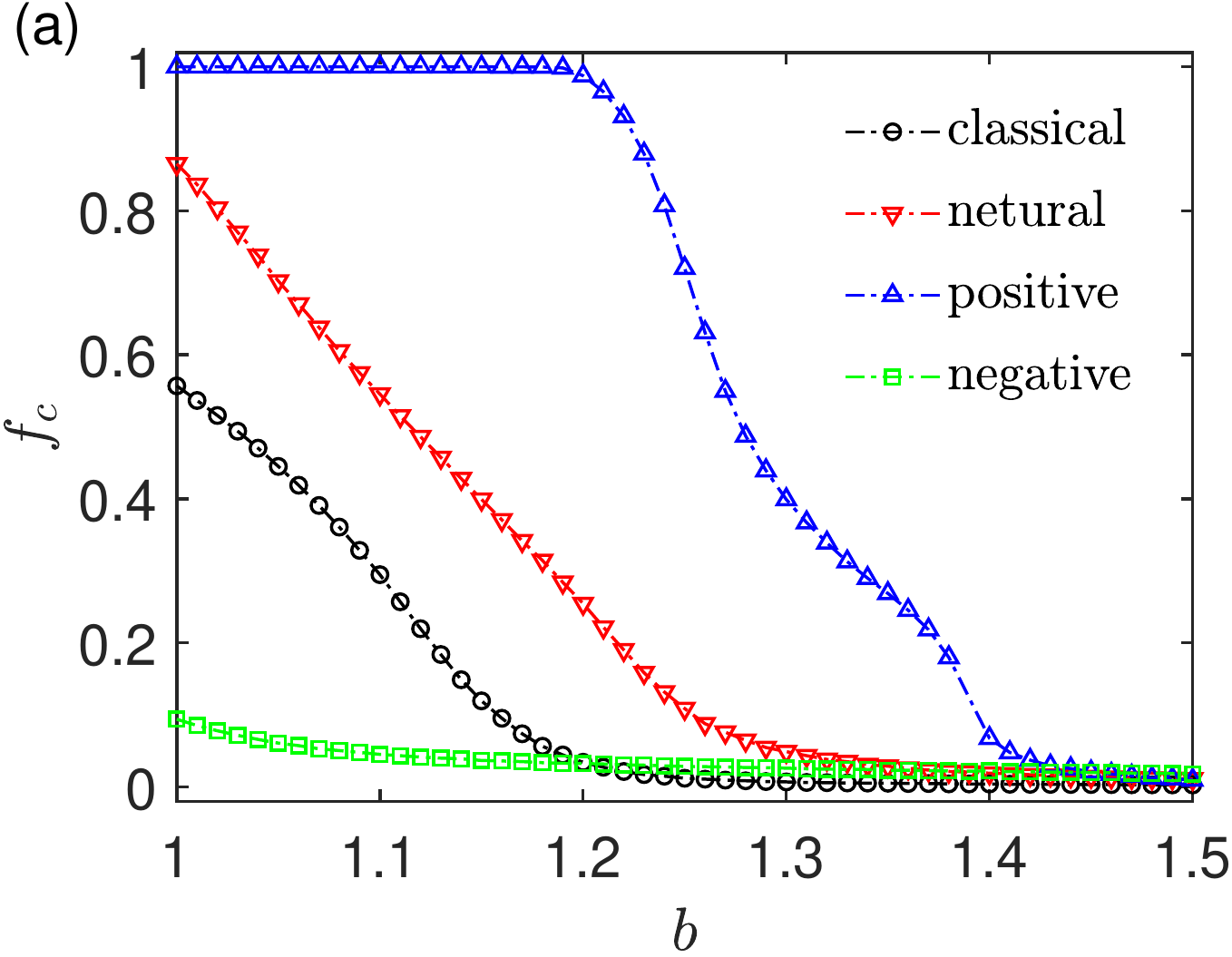}
\includegraphics[width=0.5\linewidth]{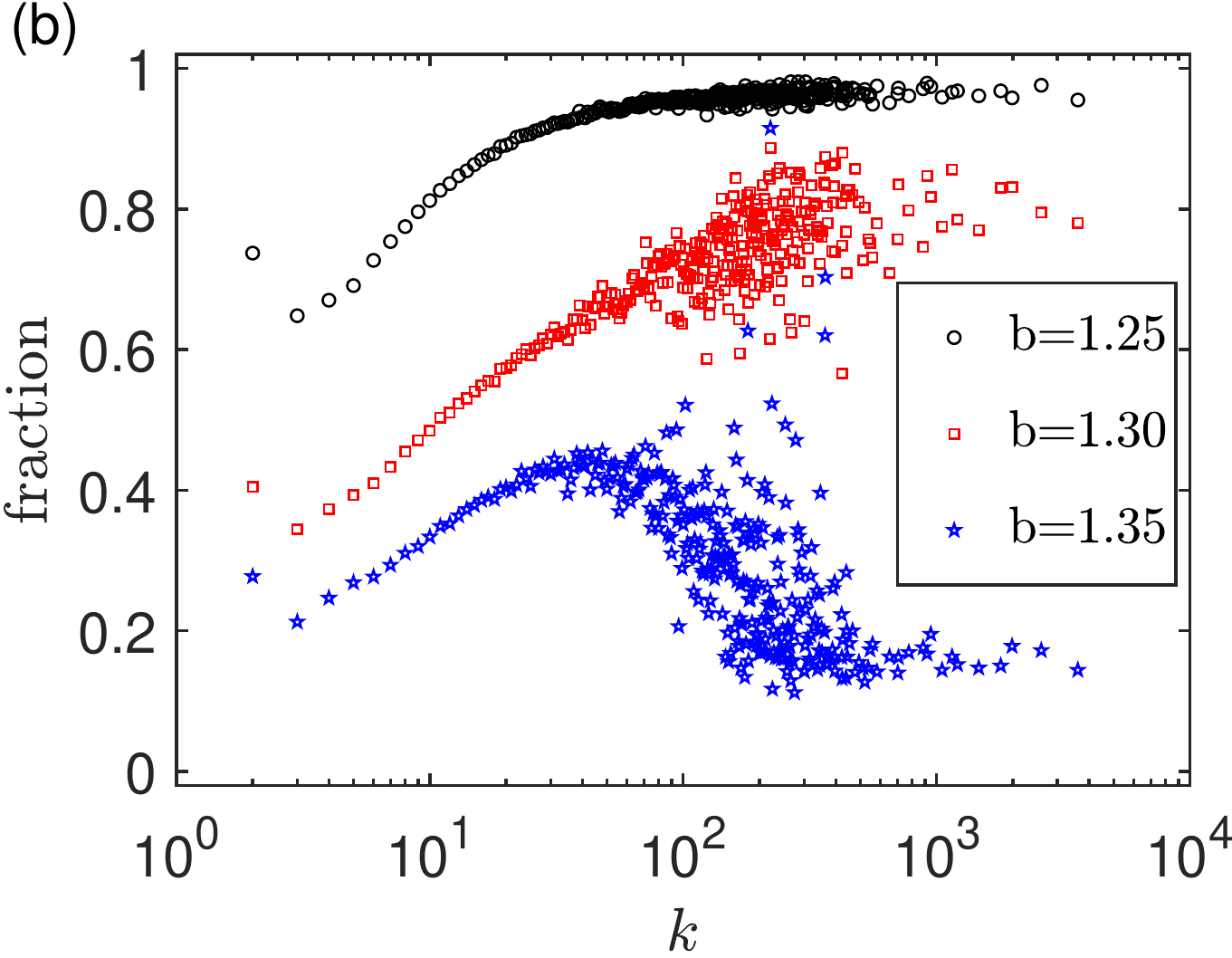}
\caption{(Color online)
Evolution on BA scale-free networks. (a) Cooperator fraction $f_c$ as a function of temptation $b$ for three different degree-hierarchy correlations (with $\alpha=1$): neutral, positive, and negative. As a comparison, the classical case ($\alpha=0$) is also shown (empty circles). (b) The fraction of cooperator versus node degrees in the networks, for three $b$ in the positive correlation case. Parameters: $N=2^{20}$, the average degree $\langle k\rangle=4$, $K=0.025$.}
\label{fig:sf_pt}
\end{figure}

\section{Hierarchical games on scale-free networks}
\label{sec:sf}
However, the square lattice studied above is a typical homogeneous network, it's natural to ask whether these findings still remain valid on other networks, like the heterogeneous graphs. To examine the validity, we also study the evolution of the hierarchical game on a scale-free network. Specifically, we adopt the Barab\'asi-Albert (BA) network model~\cite{Barabasi1999Emergence}, where the degree distribution follows a power-law with an exponent of $-3$. We also adopt the Fermi rule as Eq.(\ref{eq:imitation}). Note that the imitation probability $W(s_{j}\rightarrow s_{i})$ is based upon the mean payoffs $\bar{\Pi}_{i,j}$ as in ~\cite{Wu2007Evolutionary,Tomassini2007Social,Szolnoki2008Towards} rather than the total payoffs $\Pi_{i,j}$ as some previous work did~\cite{Santos2005Scale-Free}. Here, only the uniform hierarchy distribution is chosen. 

Due to the degree heterogeneity, the detailed implementation of hierarchy matters. In our practice, we study the following three degree-hierarchy correlations: 

(i) Neutral correlation --- the hierarchy number $h_i$ for node $i$ is randomly picked from the hierarchy distribution irrespective to its degree $k_i$;

(ii) Positive correlation --- the node with a lager degree is also of higher social hierarchy;

(iii) Negative correlation --- the node with a smaller degree is of higher social hierarchy. 

Specifically for case (ii) and (iii), firstly $N$ random number $h$ are picked from the uniform distribution, and we rank them in a descending order; the nodes are then also ranked in a descending order according to their degrees.  Positive correlation is obtained when the hierarchy number is adopted by the node at some same ranking position. When the nodes are ranked in ascending order, this then corresponds to the negative correlation implementation.   

\begin{figure}
\includegraphics[width=1\linewidth]{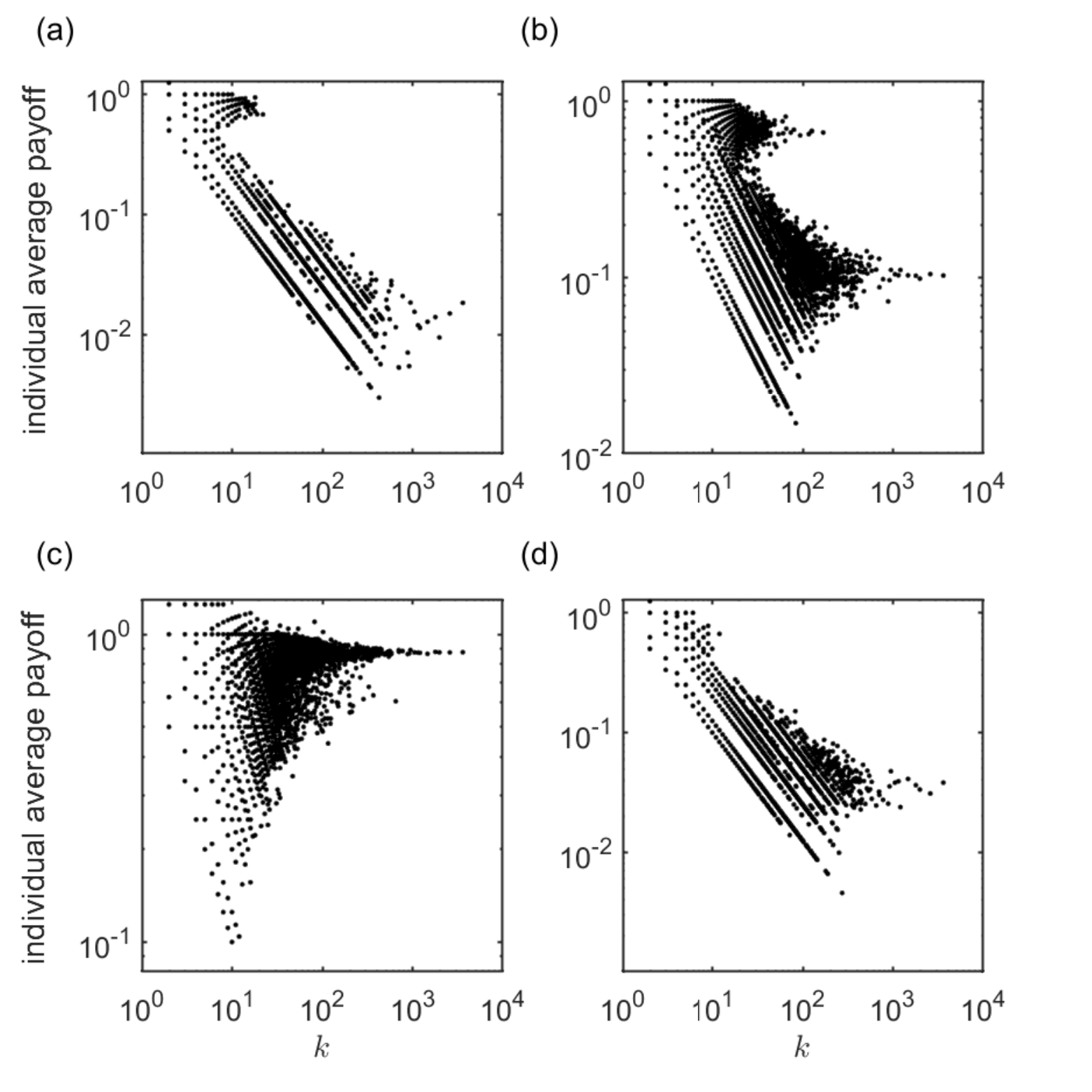}
\caption{
Scatter plots of individual average payoff versus their degree for $b=1.25$ at a steady state, for the above four cases:
(a) classical; (b) neutral; (c) positive; (d) negative. 
Other parameters: $N=2^{20}$, $K=0.025$, $\alpha=0$ for (a), $\alpha=1$ for (b-d).}
\label{fig:sf_payoff}
\end{figure}

Figure~\ref{fig:sf_pt}(a) shows the cooperation fraction for these three cases as a function of the temptation $b$. The classical scenario ($\alpha=0$) is also provided for comparison. We see that both positive and neutral correlation are able to facilitate the cooperation levels, while the negatively correlated case shows the opposite trend instead.  This observation suggests that the promotion of cooperation can also be expected in heterogeneous networks given the degree-hierarchy correlation is not of negative type. 

To understand the impact of hierarchy, it's helpful to compare the cooperation mechanism with the mean payoff $\bar{\Pi}_{i,j}$ and with the total accumulated payoff $\Pi_{i,j}$. It's now well-known~\cite{Santos2005Scale-Free,Wu2007Evolutionary,Tomassini2007Social,Szolnoki2008Towards} that in the latter case, the hubs typically have far larger total payoff than those peripheral nodes, and thus they become model players. Meanwhile, they are more likely to be cooperators because defection is unstable in the long term. As a result, the hubs act the cooperation backbone of the population, driving the overall cooperation to a high level. By contrast, when using the mean payoff as did in our work, the payoff advantage of hubs is much reduced (see Fig.~\ref{fig:sf_payoff}(a)), they are no longer the model players. Furthermore, they are not likely to be cooperators, see e.g the case of $b=1.35$ in Fig.~\ref{fig:sf_pt}(b). In fact, the whole network is separated by these hubs into some small cooperator fragments. There is no cooperation backbone and the cooperation is much lower than the other case. The cooperation is inhibited by the degree normalization.

Now in the positive correlation case, where hubs are endowed with higher social ranks, hubs again becomes more likely to be model players due to the hierarchical impact, therefore the presence of hierarchy recovers their advantages as in the total payoff case (see Fig.~\ref{fig:sf_payoff}(c)). As a consequence, the hubs are more likely to be cooperators in the longer run (see $b=1.25$ and $1.30$ in Fig.~\ref{fig:sf_pt}(b)). These hubs form a cooperation backbone that drives the whole population to a higher cooperation level. Instead, when the negative correlation is posed, the disadvantage of hubs are further enhanced (see Fig.~\ref{fig:sf_payoff}(d)), the cooperation inhibition as in the mean payoff case becomes more apparent, thus a lower cooperation level is expected as shown in Fig.~\ref{fig:sf_pt}(a). Fig.~\ref{fig:sf_payoff}(b) shows that the neutral  correlation case exhibits two peaks for individual average payoffs, somewhere between the positive and negative case. But the payoffs for hubs are approximately one order larger than the classical case, this may explain that the cooperation in neutral case is better than the classical case.

In fact, we made qualitatively the same observations when using total accumulated payoffs (data not shown), where the positive correlation leads to cooperation enhancement, negative correlation to a decrease, and neutral case brings no obvious change.  But since the cooperation prevalence in the baseline case ($\alpha=0$) is already fairly high, the enhancement in the positive case is not so significant as in Fig.~\ref{fig:sf_pt}(a). The overall picture is that the advantage of hub is enhanced in the positive correlation case, while it is inhibited in the negative case. Detailed analysis further shows that when using accumulated payoffs the hubs always form stable cooperation backbone of the population, different hierarchy implementations only alter the cooperator fraction in the periphery nodes. 

\begin{figure}
\includegraphics[width=1\linewidth]{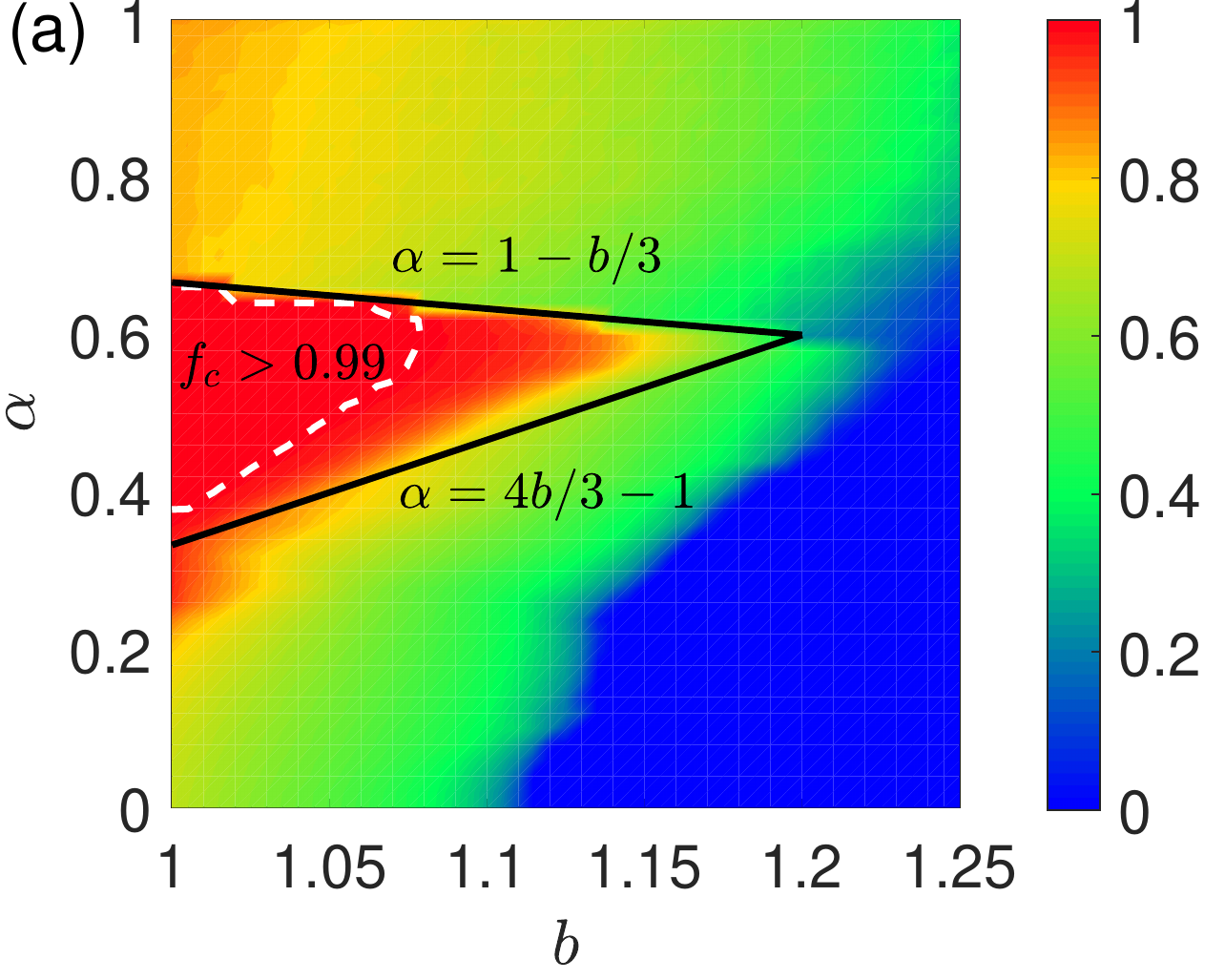}
\includegraphics[width=1\linewidth]{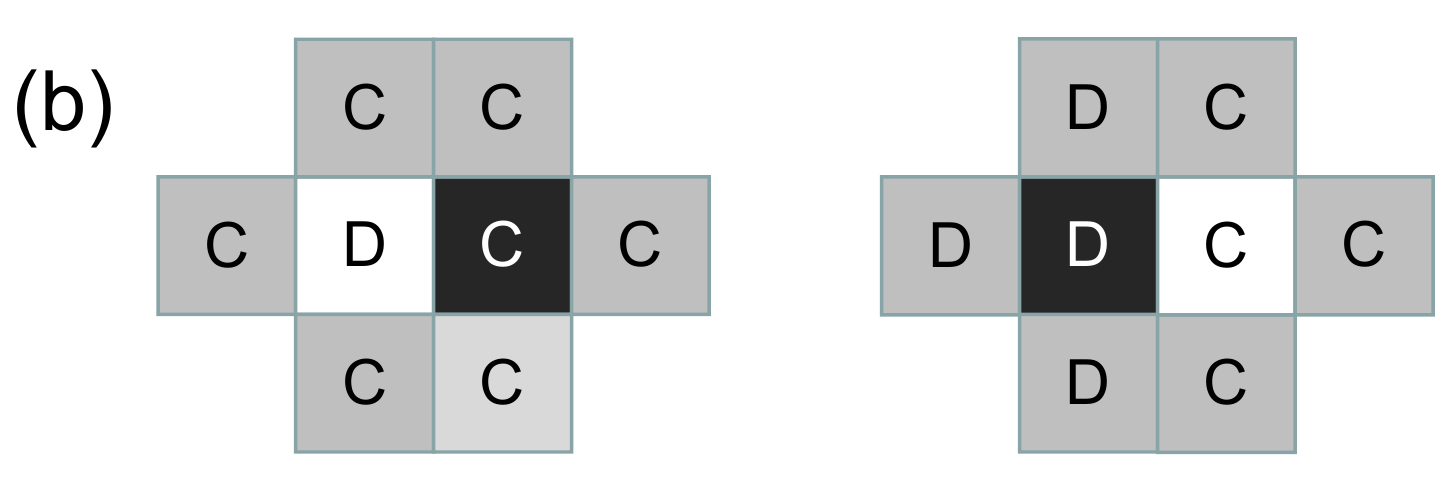}
\caption{(Color online)
A two-hierarchy model. (a) Color coded fraction of cooperation $f_c$ in the parameter $b$-$\alpha$ space. In particular, the nearly full cooperation ($f_c>0.99$) is encircled by white dashed line. The system evolves 510000 MC steps and the data is the average of the last 10000 points. Parameters: $L=256$, $K=0.025$.
(b) Two microscopic schemes for the derivation of full cooperation conditions:
(Left) a low rank defector is surrounded by cooperators with one of them being of a high rank;
(Right) a cluster of defectors encounter a cluster of cooperators, where the focal defector/cooperator is of high/low rank. White and black color indicate the social rank being 0 and 1, respectively, and grey sites could be in either rank.
}
\label{fig:simplified_model}
\end{figure}

\section{Some analytical treatments}
\label{sec:analysis}
Finally, to gain some analytical insight into the hierarchy impact, we adopt a further simplified model on 2d-square lattice, where only two hierarchies are assumed within the population (i.e. $h_i\in\{0,1\}$), and the replicator rule is used. The probability of strategy adoption is as follows:
\begin{equation}
 \begin{aligned}
 W(s_{j}\rightarrow s_{i})=\max\{\dfrac{\bar{\Pi}_{j} (1+\alpha\bigtriangleup h)-\bar{\Pi}_{i}}{b(1+\alpha)},0\},
 \end{aligned}
\end{equation}
where $\bar{\Pi}_{i}$ and $\bar{\Pi}_{j}$ are the mean payoffs of $i$ and $j$ as above. In this two-hierarchy model, $\bigtriangleup h=h_{j}-h_{i}$ has only three values: 0, $\pm1$. When the effective payoff $\Pi_{j} (1+\alpha\bigtriangleup h)>\Pi_{i}$, the player $i$ adopts player $j$'s strategy with a nonzero probability, otherwise there is no change in $s_i$. The presence of $b(1+\alpha)$ is for the probability normalization. The reason for the adoption of the replicator rule is because it is more readily for analytical treatment than the Fermi rule.

Fig.~\ref{fig:simplified_model}(a) shows the results of numerical experiments with this simplified model, which exhibit qualitatively the same behaviors that the presence of hierarchy is able to enhance cooperation prevalence. In particular, the defector eradication threshold $b_{c_2}$ is also shown to be a linear function of the hierarchical parameter $\alpha$. Though new complexities are revealed that an upper threshold of full cooperation arises.

To have a stable full cooperation state, one can consider an extreme case where a single defector as a perturbation is surrounded by a group of cooperators, and find out under what condition this defector is going to die or at least there is a possibility for it to be invaded by cooperators of any hierarchy, and thus reaching absorbing state of full cooperation is possible. When this defector is of low rank ($h_D=0$), a necessary condition is that if one of its cooperator neighbors is of high rank ($h_C=1$) and the state transfer probability requires $W(C\rightarrow D)>0$ (shown in the left panel of Fig.~\ref{fig:simplified_model}(b)). This scheme corresponds to the loosest scenario for defector extinction. The effective payoffs are $3(1+\alpha)$ and $4b$ respectively for the focal cooperator and defector. This leads to the following inequality
 \begin{equation}
 \alpha>4b/3-1.
 \label{eq:below}
\end{equation}

A tough situation occurs when the defector is of high rank ($h_D=1$). Since its effective payoff is higher than its any cooperator neighbor, it will convince some of its neighbors to be defectors after a few steps. To become full cooperation, this requires this defector cluster can be invaded. The most difficult scenario in this case is shown in the right panel of Fig.~\ref{fig:simplified_model}(b), and also $W(C\rightarrow D)>0$ is required. Here the effective payoffs of the two focal players are respectively $3(1-\alpha)$ and $b$ for C and D, and we have
 \begin{equation}
 \alpha<1-b/3.
\label{eq:upper}
\end{equation}

The equations of the inequality (\ref{eq:below}) and (\ref{eq:upper}) constitute the boundaries of the full cooperation region, which are well matched by numerical results, see the black lines in Fig.~\ref{fig:simplified_model}(a). But since the analytic boundaries are derived from the necessary conditions, the two boundaries only encircle the full cooperation region, cannot reproduce its exact boundary.

\section{Summary}
\label{sec:summary}
In summary, we show that the social hierarchy, as a ubiquitous observation in nature and human society, can effectively promote the cooperation outcome in structured population with all the three hierarchy distributions considered. 
The mechanism for cooperation promotion lies in the fact that the hierarchy in the population facilitates the formation of cooperation clusters that effectively resist the invasion of defection. We argue that this promotion effect is conceptually similar to heterogeneity-induced promotion in some previous diversity studies, such as the structural heterogeneity~\cite{Santos2005Scale-Free,Santos2008Social} or in~\cite{Szolnoki2007Cooperation,Perce2008Social}.
Detailed comparison shows that the uniform distribution works the best while the power-law the worst instead,
which is opposite compared to previous diversity studies~\cite{Perce2008Social,Qin2017Social}.
These findings are not limited to lattice systems, they remain valid also on BA scale-free networks. 
In particular, when a positive degree-hierarchy correlation is posed, the hubs' advantage is enhanced to the largest degree. Interestingly, the adopted mean payoff scheme implies an inhibited structural heterogeneity, the introduced hierarchy is instead to partially restore the impact of heterogeneity, compensate the inhibition effect.

We would like to stress that although the evolution outcome and mechanism in our study is akin to some previous diversity studies, this work focuses on the role of hierarchy, i.e. explicitly the impact of the feature difference between individuals rather than the feature per se.
Our findings may provide an plausible explanation of the ubiquitousness of social hierarchies and implies that introducing some degree of hierarchy into the population seems an optional strategy for institutional design to boost cooperation. Besides, our work calls for behavioral experiments for further confirm.

\section{Acknowledgments}
We are supported by the Natural Science Foundation of China under Grants No. 61703257 and 12075144, and by the Fundamental Research Funds for the Central Universities GK201903012.

\bibliographystyle{elsarticle-num-names} 
\bibliography{hierarchical}









\end{document}